\documentclass[12pt]{article}

\usepackage{scicite}
\usepackage{graphicx}

\usepackage{times}

\newenvironment{sciabstract}{%
\begin{quote} \bf}
{\end{quote}}

\newcounter{lastnote}

\title{Nucleosome shape dictates chromatin-fiber structure}

\author{Martin Depken,$^{1,2,3\ast}$ Helmut Schiessel$^{3}$\\
\normalsize{$^{1}$ Max Planck Institute for the Physics Complex Systems,}\\
\normalsize{N\"othnitzer Stra\ss e 38, 01187 Dresden, Germany}\\
\normalsize{$^{2}$Max-Planck Institute for Molecular Cell Biology and Genetics,}\\
\normalsize{Pfotenhauerstr. 108, 01307 Dresden, Germany}\\
\normalsize{$^{3}$Instituut-Lorentz for Theoretical Physics, Universiteit Leiden,}\\
\normalsize{ Postbus 9506, 2300 RA Leiden, The Netherlands}\\
\normalsize{$^\ast$To whom correspondence should be addressed; E-mail:  depken@pks.mpg.de.}
}

\date{}

\begin{document} 




\maketitle


\begin{sciabstract}
Apart from being the gateway for all access to the eukaryotic genome, chromatin has in recent years been identified as carrying an epigenetic code regulating transcriptional activity. The detailed knowledge of this code contrasts the ignorance of the fiber structure which it regulates, and none of the suggested fiber models are capable of predicting the most basic quantities of the fiber (diameter, nucleosome line density, etc.). We address this three-decade-old problem by constructing a simple geometrical model based  on the nucleosome shape alone.  Without fit parameters we predict the observed properties of the condensed chromatin fiber (e.g.  its 30 nm diameter), the structure, and how the fiber changes with varying nucleosome repeat length. Our approach further puts the plethora of previously suggested models within a coherent framework, and opens the door to detailed studies of the interplay between chromatin structure and function. 
\end{sciabstract}


Eukaryotic DNA is wrapped around histone proteins, resulting in a string of wedge-shaped nucleosomes connected via short stretches of linker DNA. Under physiological salt concentrations this string can undergo additional folding, forming what is referred to as the 30 nm fiber. Whereas the structure of the nucleosome is known to atomic resolution\cite{Luger1997}, the three-dimensional arrangement of nucleosomes in the 30 nm fiber remains poorly understood --- despite three decades of experiments and model building\cite{Holde2007}. The wide range of models documented in the literature can be divided into  roughly two classes: the traditional solenoid models\cite{Finch1976} and the crossed linker models\cite{Woodcock1993}. Unfortunately, neither type provide criteria to identify the optimal fiber geometry, and basic quantities like the fiber diameter are ultimately fixed by the fine tuning of unknown parameters. This lack of predictive power and the fact that many experiments\cite{Widom1985, Williams1986, Worcel1981, McGhee1983, Sen1986}  were likely performed on amorphous samples\cite{Holde1995} are arguably  the main reasons why the structure has remained unresolved for so long. Some of the experimental issues have recently been overcome through the reconstitution of highly regular fibers\cite{Robinson2006,Dorigo2004}, but a comprehensive modeling effort determining the structure is still lacking. We take inspiration from previous success of geometric arguments when applied to information-carrying structures\cite{watson1953} and address this problem by considering all fibers satisfying the geometric condition that the nucleosome core particles (NCPs) pack densely on the periphery of the fiber\cite{Holde2007} (see Fig.~1 a). Without free parameters this enables us to make definite predictions that are all born out when compared to the new regular reconstituted fibers\cite{Robinson2006,Dorigo2004}.  Our approach makes the implicit assumption the short range attraction between NCPs\cite{Mangenot2002, Mangenot2002a, Muehlbacher2006} constitute the dominant mode of interaction in dense chromatin fibers, being only weakly modulated by the soft contribution from the DNA-linker backbone. Taking this view, the problem of determining the structure of the chromatin fiber splits into two parts: the identification of dense configurations of NCPs on the periphery of the fiber, and the estimation of the energetic contribution from the linker backbone in order to determine which structure is realized.

First addressing the formation of a dense shell we note that NCPs aggregate into arcs in solution\cite{Dubochet1978}, indicating that their
wedge-shaped form\cite{Luger1997} can play a key role in dictating
large scale arrangements of interacting nucleosomes. Drawing on
this, we take the {\em effective shape} of the NCP as being that of
a wedge shaped cylinder (see Fig.~1 b). In Figure~1 c we show how a dense packing of nucleosome
footprints on a periodic strip is a necessary condition for a
dense three-dimensional packing of nucleosomes in the fiber.   By assuming a dense packing the footprints are forced into helical ribbons winding along the fiber. These are formed along either of the footprints
symmetry axes. Models normally referred to as
interdigitated\cite{Robinson2006,Robinson2006a} belong to the set
of dense packings where the ribbons form along the major axis
(NCPs stacking side to side). Here we assume the NCPs to stack
face to face, corresponding to footprints forming ribbons along
their minor axis (see Supplementary note for the modifications to the below in the case of interdigitated structures). Such stackings of NCPs aggregate spontaneously
under the right solvent
conditions\cite{Dubochet1978,Mangenot2003}, and is also what best
utilizes any short range attractive interaction. For a dense
footprint packing, the nucleosome line density (NLD) $\sigma$ is
simply the width of the strip onto which they pack, divided by the
foot print area,
\begin{equation}
\sigma=\frac{\pi(D-a)}{ab}.
\end{equation}
The manner in which the  ribbons spiral up along the
fiber, parameterized by the ribbon angle $\gamma$, is set by the
requirement that the $N_{\rm rib}$ ribbons precisely fill up the periodic strip
(see Fig.~1 c),
 \begin{equation}
 N_{\rm rib} \frac{a}{\cos \gamma}=\pi (D-a).
 \end{equation}
In addition we require that the backbone connects all the
nucleosomes in a regular fashion. Denote by $N_{\rm step}$ the distance
across ribbons between connected nucleosomes (see
Fig.~1 a). The necessary and sufficient condition
for a regular backbone winding (BW) --- completely defined by
the pair $(N_{\rm rib},N_{\rm step})$ --- is the existence of two
integers $n$ and $k$ with $0\le n\le k\le N_{\rm rib}$ such that 
\begin{equation}
\label{wind} k N_{\rm step}-n N_{\rm rib}=1.
\end{equation}
Equation 3 ensures that neighboring ribbons are
eventually connected (after $k$ steps and $n$ turns, see Supplementary note) and hence all ribbons are visited by the backbone (c.f. $(4,2)$, which only connects half of the NCPs). The trivial BW
$(N_{\rm rib},1)$ corresponds to the backbone connecting
nucleosomes in neighboring ribbons (since $N_{\rm step}=1$, see
Fig.~1 a). Such a backbone can be found for fibers
with any number of ribbons since with $n=0$ and $k=1$ condition Equation
3 is always satisfied. The classical solenoid
model\cite{Finch1976} has a $(1,1)$ BW, and all the models
considered by Wong {\em et al.}\cite{Wong2007} have  trivial
BWs. By scanning through the finite number of possible $n$'s and
$k$'s one finds all additional non-trivial BWs, extending the set of 
crossed-linker models to $(5,2)$, $(7,2)$, $(7,3)$, $(8,3)$, and
so on.  Thus this approach exhaustively covers all major contending models for the fiber structure\cite{Finch1976,Worcel1981,Makarov1985,Williams1986,Woodcock1993,Holde2007}  (solenoid models, crossed linker models, interdigitated models, etc.), including some specific models not previously considered, and puts them firmly within a coherent framework.

Returning to the full three-dimensional packing of nucleosomes we
note that all admissible footprint packings correspond to NCPs
packed together with different {\em effective} wedge angles (see Fig. 1 b). 
Through considering the curvature
along the ribbons it is straightforward to relate the effective
wedge angle to the fiber diameter (see Supplementary note). Within the relevant
parameter ranges the exact expression can be approximated as
\begin{equation}
\alpha\approx\frac{2 b
}{D-a } \left(1-\left[\frac{a N_{\rm rib}}{\pi (D-a)}\right]^2\right), \label{alpha}
\end{equation}
with an accuracy of a couple of tenths of a degree. For any
specific fiber diameter $D$ and number of ribbons
$N_{\rm rib}$ this directly gives the effective wedge angle, and
it can easily be inverted to give the possible fiber diameters for
any specific effective wedge angle. In Figure~2 a
we show how the effective wedge angle varies with fiber diameter
for fibers with up to ten ribbons. In what follows we will use the
effective NCP diameter $a=11.5$ nm and average height $b=6.0$ nm
as deduced for the close packings of NCPs into columnar
quasi-hexagonal  crystals\cite{Mangenot2003} under physiological
salt concentrations and moderate pressures. We are ultimately
interested in the {\em in vivo} situation where there are
additional linker histones present, bringing the in and out going
DNA at each nucleosome into a stem
structure\cite{Bednar1998} (see Fig.~1 b). Taking
this into account, we require  $D>2(l_{\rm stem}+a)$ in
order to avoid steric interactions between stems on
opposite sides of the fiber. Here $l_{\rm stem}=3$ nm  is the
length of the induced stem as measured by Bednar {\em et al.}\cite{Bednar1998}. In
Figure~2 b we illustrate all fibers and BWs with a
diameter of 33 nm. They include the solenoid model
$(1,1)$\cite{Finch1976}, the two-start helix
$(2,1)$\cite{Dorigo2004} and the crossed linker model
$(5,2)$\cite{Woodcock1993}. It is clear from the number of
possible structures that fixing the fiber diameter tells us little
about the internal structure of the fiber, though it explains the
wide range of models suggested in the literature; all made
consistent with the experimental findings but with little
predictive power. Instead we take a reductionist approach and enforce the microscopic condition of optimal dense face-to-face stacking of nucleosomes. In  experiments by Dubochet and Noll\cite{Dubochet1978} unconstrained nucleosomal arcs were observed  with the effective wedge angle $\alpha=8^\circ$ for the NCP repeat unit. This will be the effective
wedge angle assumed throughout the rest of this paper.  With this
microscopic condition we directly get a discrete set of possible
shell structures, three of which are shown in
Figure~2 c (structures A, B, and C), and all of which are clearly
distinguished from each other on the level of the fiber diameter
and NLD (see Table~1). Here we do not discuss
the very wide fibers, one of which is displayed in the
inset of Figure~2~a. These structures might never
be realized in chromatin, but are similar to the gigantic tubes of
NCPs observed by Dubochet and Noll\cite{Dubochet1978}. The results
of the simple assumption of a dense packing of nucleosomal wedges
are summarized in Table~1, where we list all
fibers with a fiber diameter up to 63 nm. As detailed below some of these have already been observed, while others might still be found through further experiments.

Armed with a small set of possible shell structures, we now
examine the linker backbone to determine which of these is
realized for any specific nucleosomal repeat length.  Though we lack a precise model for the energetics of the backbone, we can still
put upper and lower bounds on the possible linker lengths for a
specific shell structure and BW. The lower bound is set by the
shortest distance between two successive stems along the backbone.
This depends not only on the BW $(N_{\rm rib},N_{\rm step})$ but
also on the relative helicity of backbone and ribbons. We denote
structures where ribbons and backbone have the same helicity by
$(N_{\rm rib},N_{\rm step})^+$, and by $(N_{\rm rib},N_{\rm
step})^-$ in the opposite case. The upper limit  for the linker
length is set by the excluded volume constraint on the inside of
the fiber. We assume that due to the presence of cationic histone
tails the highly charged linker DNA can be hexagonally packed with a shortest centre-to-centre distance set by the DNA diameter
$d_{\rm DNA}=2$ nm. The resulting limits on the linker lengths are
indicated in Table~1. We thus see that for the
shortest repeat lengths the realized structure must always be
$(5,1)^\pm$ or $(5,2)^{\pm }$. These feature a 33 nm diameter,
from which the 30 nm fiber derives its name. Of these two
structures we expect $(5,2)^{\pm}$ to be realized since it allows for
the straightest linkers (see Fig.~2 c, structure
A). When increasing the linker length the fiber must take on
another structure before the maximum repeat length of 210 bp. The
fact  that $(7,3)^\pm$ (possible for repeat lengths over 207 bp)
has the straightest linkers (see Fig.~2 c,
structure C) makes it a good candidate for the target
structure. 
Thus, through the very simple, geometric, and microscopic condition of an optimal nucleosome packing combined with rudimentary arguments concerning the backbone, we are able to make predictions concerning the precise structure realized for different nucleosomal repeat lengths.

Having discussed the theoretically possible fibers for different repeat lengths, and identified a plausible transition point between structures, we now compare this with recent results on dense reconstituted fibers. Robinson {\em et al.}\cite{Robinson2006} observed that such fibers clustered into two sets, each signified by a specific fiber diameter and NLD  (see Fig.~3). As pointed out by Wong {\em et al.}\cite{Wong2007}, a similar clustering is also seen for the native fibers examined by Williams {\em et al.}\cite{Williams1986}. Since the fiber diameter and the NLD are linearly related through condition~Equation 1, we can use this as a direct test of our approach. In Figure~3 we plot this relation together with the observed fiber diameter and nucleosome line densities\cite{Robinson2006}. Our model is consistent with the experimental data and manages to account for both thin and thick fibers without any fit parameters. In Figure~2 we have indicated the average diameters of the two clusters observed by Robinson {\em et al.}\cite{Robinson2006}. The structures predicted by our model (Fig.~2 a and c, structures A and C) are the only fibers within the error bars of the experiments.  It can also be seen that these predictions are rather robust against changes in the effective wedge angle. In addition, the transition between fiber structures  is observed  somewhere between repeat lengths of 207 and 217 bp, which is also captured by our model. Apparently contradicting these results is another recent set of experiments\cite{Dorigo2004} suggesting a two-ribbon structure.
These were performed on short fibers (10-12 nucleosomes), and would thus be unlikely to capture the structures suggested here (5 and 7 ribbons) since they only allow for around two nucleosomes to stack in each ribbon. In line with the basic assumption of our model, that nucleosome interactions drive the assembly of the fiber, we expect short fibers to favor fewer ribbons in order to minimize the number of nucleosome faces exposed to the solution. The same reasoning applies to inferring the arrangement of nucleosomes in the fiber from the crystallographic structure of the tetra-nucleosome\cite{Schalch2005}. Thus we conclude that our zero-fit-parameter model is consistent with all experimental findings on regular fibers to date.

Our model could be further tested through e.g. linear
dichroism\cite{McGhee1983,Sen1986} studies determining the ribbon
angle $\gamma$ for these newly available  structures. It would
also be interesting to see if structures like $(8,3)^\pm$ or
$(9,4)^\pm$ are ever realized for repeat lengths  longer than
those investigated by Robinson {\em et al.}\cite{Robinson2006}. Moving
away from the 10 bp ladder (set by DNA's helical repeat length)
used by Robinson {\em et al.} would further elucidate whether  it is the bend or twist energy of the linker DNA that dictates the chosen structures. 
Our model also suggests that the predicted structures are insensitive to a certain amount of variations in the nucleosome repeat length. With the inclusion of a energetic model for the linker backbone the above development should form the basis for statistical and kinetic  studies of
how {\em in vivo} variations and correlation in repeat
length\cite{Segal2006} affect the locally realized structure,  its stability, and thus the observed condensation-decondensation transition of the
30 nm fiber. {\it In vitro}, this transition can be probed
by a change of ionic conditions\cite{Bednar1998}, or the application of a
sufficiently large external force, e.g. in a
single-molecule-experiment using optical tweezers\cite{Cui2000}. {\it In vivo},
this can be done by the acetylation of histone tails\cite{Horn2002,Dion2005}, thus
offering a straightforward way of increasing the accessibility to
the packed genetic material. Understanding the structure of the fiber now opens the door to detailed study of this transition, and its connection to the histone code\cite{Strahl2000,Jenuwein2001,Dion2005}. Ultimately this structural knowledge should be combined with biochemical studies in order to move towards a comprehensive understanding of the subtle interplay between structure and function in chromatin\cite{Kornberg2007}.
\subsection*{Acknowledgments}
We would like to acknowledge fruitful discussions with Ralf Everaers, Nima Hamedani Radja, Marc Emanuel, and John van Noort. We thank Eric Galburt, Stephan Grill, and Mirjam Mayer for comments on the manuscript. A special thank is owed Wim van Saarloos without whom this research would not have been
 possible. M.D. acknowledges support from the physics foundation
 FOM.

\bibliography{/Users/depken/Desktop/Documents/Work/ReferencesBIOPHYS/refs}
\bibliographystyle{Science}


\pagebreak

\begin{figure}
\begin{center}\includegraphics[width=.6\textwidth]{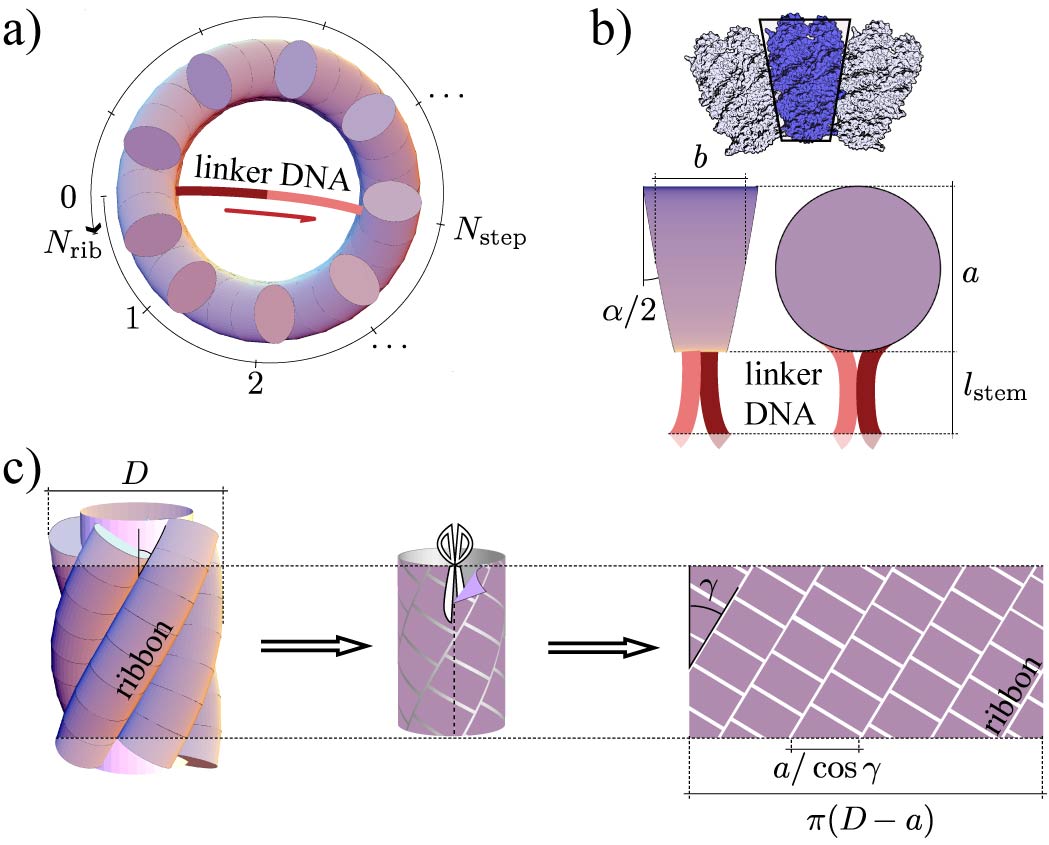}\end{center}
\caption{The model of the dense chromatin fiber. {\bf a)} In a dense chromatin fiber the nucleosomes  pack on the
outside of the fiber, and the linkers are situated on the inside. Here the
$(N_{\rm rib}=9,N_{\rm step}=4)$ backbone is illustrated, completely
specifying the way the nucleosomes are connected. {\bf 
b)} Illustration of how three nucleosomes stack, and how this relates to
the  {\em effective} wedge shape of the nucleosome (outlined). By using the term
effective wedge shape we stress that the nucleosome itself need
not form a perfect wedge, but rather that we
rely on the experimental observation\cite{Dubochet1978} that when
they aggregate one can identify a wedge shaped repeat unit. Onto
the wedge shaped cylinder we attach a rigid stem\cite{Bednar1998} to represent the
in and out going DNA and a linker histone (not shown). {\bf c)}  Illustration of how a dense nucleosome-footprint packing on a cylinder running through the nucleosome centers is a necessary condition for a dense  three-dimensional structure. Also indicated are the ribbons induced by the dense packing, together with the angle $\gamma$ they make with the fiber axis. \label{fig:rollout} }
\end{figure}

\begin{figure}
\begin{center}\includegraphics[width=.5\textwidth]{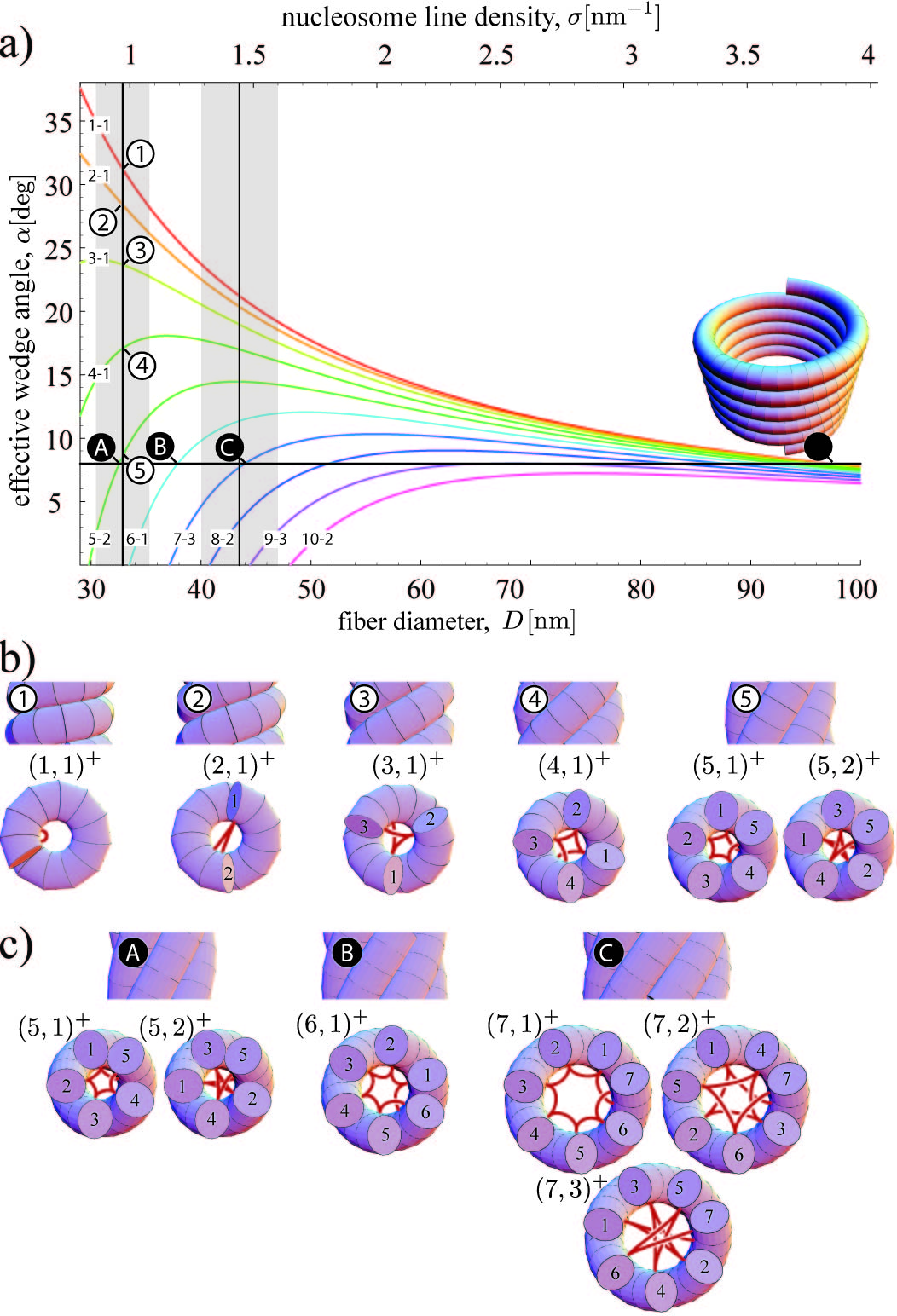}\end{center}
\caption{The predicted fibers. {\bf a)} Plot of how the effective wedge angle $\alpha$ varies with fiber diameter $D$ (or, equivalently nucleosome line density $\sigma$) for fibers with up to ten ribbons. Each curve is labeled as: number of ribbons - number of possible backbones. Indicated is also the effective wedge angle measured by Dubochet and Noll\cite{Dubochet1978} and used in this paper to predict the
fiber structures. A giant fiber solution with 87 nm diameter, and similar to the ones  seen by Dubochet and Noll\cite{Dubochet1978}, is displayed in the inset. Further indicated are the average fiber diameters for the two different sets of fibers observed by Robinson {\em et al.}\cite{Robinson2006} (see also Fig.~3), with the area including one standard deviation indicated in grey. {\bf b)} All solutions with different effective wedge angles achieved by fixing the fiber diameter to 33 nm (point 1-5 in Panel a), together with the available backbones. {\bf c)} Some of the structures predicted by fixing the effective wedge angle to the value measured  by Dubochet and Noll\cite{Dubochet1978} (points A-C in Panel a), together with the allowed backbones. The fibers with 5 (A) and 7 (C) ribbons come very close to the structures observed by Robinson {\em et al.}\cite{Robinson2006} for dense reconstituted fibers. \label{fig:solutions}}
\end{figure}

\begin{figure}
\begin{center}
\includegraphics[width=.6\textwidth]{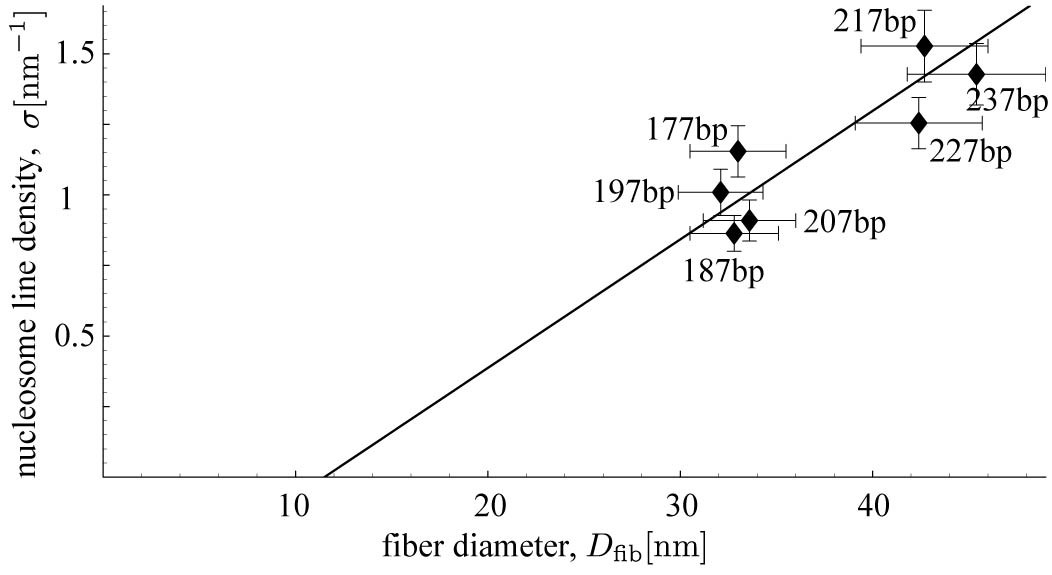}\end{center}
\caption{Experimental observations on dense fibers. Re-representation of data published by Robinson {\em et al.}\cite{Robinson2006}
for  reconstituted fibers with  different nucleosomal repeat
lengths (indicated). The data points are seen to cluster around two
specific diameters, $D=$33 nm and 44 nm, and nucleosome line densities, $\sigma=$1.0 nm$^{-1}$ and 1.4 nm$^{-1}$. The error bars indicate one standard deviation. Also
shown is the linear relation between fiber
diameter and nucleosome line density predicted by our model~Eq. 1. It is seen to be
consistent with both thin and thick fibers, without any adjustable parameters.\label{fig:scatter}}
\end{figure}

\begin{table}
\caption{The predicted fiber structures. The table displays all the calculated properties of the
fibers consistent with the structure of the nucleosome. They are (left to right): fiber structure, fiber diameter, nucleosome line density, angle formed between ribbons and fiber axis, minimum nucleosomal repeat length, and maximum nucleosomal repeat length. The minimum
repeat length differs for different relative helicities between
linker backbone and ribbons, the value for opposite relative
helicity being indicated within parenthesis (for the case of at
least one bp difference between the two). Empty fields take the values of fields
directly above.
  }
\begin{tabular}{ccccccc}
~\\
Structure                   &$D$(nm)      &$\sigma$(1/nm)   &$|\gamma|(deg)$ &$l_{\rm repeat}(bp)>$  &$l_{\rm repeat}(bp)<$  \cr
\hline
$(5,1)^\pm$             &33               &1.0     &29    &172(171)     &210          \cr
$(5,2)^\pm$             &                   &               &                   &175          &                   \cr
 $(6,1)^\pm$        &38               &1.2     &33         &178(177)     &269          \cr
$(7,1)^\pm$             &44               &1.5     &38         &184(183)     &343          \cr
$(7,2)^\pm$             &                   &           &               &199(198)     &               \cr
$(7,3)^\pm$             &                   &           &               &207          &               \cr
$(8,1)^\pm$             &52               &1.8     &42         &190(189)     &440          \cr
$(8,3)^\pm$         &                   &           &               &225          &               \cr
$(9,1)^+$           &63               &2.4     &50         &200(198)     &605          \cr
$(9,2)^+$           &                   &           &               &230(229)     &               \cr
$(9,4)^\pm$             &                   &           &               &264          &               \cr
\hline
\end{tabular}
 \end{table}
~
\newpage
~
\pagebreak
\section*{Supporting Notes}
\renewcommand{\thefigure}{S\arabic{figure}}
\setcounter{figure}{0}

\paragraph*{The relation defining the backbone winding:}
\begin{figure}[htp!]
\begin{center}\includegraphics[width=.44\textwidth]{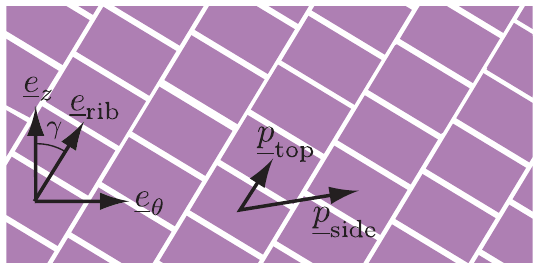}\end{center}~\\[-1.5cm]
\caption{Footprint packing structure. The footprint packing defined in Figure~1 c of the main text, with lattice vectors $\underline{p}_{\rm top}$ and $\underline{p}_{\rm side}$ indicated, together with the unit vectors $\underline{e}_{z}$, $\underline{e}_{r}$, and $\underline{e}_{\rm rib}$. \label{fig:s1}}
\end{figure}
Building on the footprint packings displayed in Figure 1 of the main text, define the vector $\underline{p}_{\rm top}$ as the vector connecting the nucleosome at its base (see Fig.~S1) with the next nucleosome placed in the same ribbon which is encountered when moving along the backbone. In a similar manner define $\underline{p}_{\rm side}$ as ending at the next nucleosome encountered in the ribbon neighboring to the right. Let $k_{\rm top}$ be the number of steps taken along the backbone when going between the nucleosomes connected by  $\underline{p}_{\rm top}$, and $n_{\rm top}$ the number of times  the fiber was circled in doing so. Define $k_{\rm side}$ and $n_{\rm side}$ in the analogous manner. Then we have
$$
\underline{p}_{\rm top}=(k_{\rm top}\Delta -\pi (D-a) n_{\rm top})\underline{e}_{\theta}+\frac{k_{\rm top}}{\sigma}\underline{e}_{ z},
$$
$$
\underline{p}_{\rm side}=(k_{\rm top}\Delta -\pi (D-a) n_{\rm side})\underline{e}_{\theta}+\frac{k_{\rm side}}{\sigma}\underline{e}_{ z},
$$
where $\sigma$ is the nucleosome line density along the fiber, and $\Delta$ is the circumferential distance between nucleosomes following each other along the backbone.
For the packing of ribbons to be dense, the parallelogram spanned by these two vectors must be of the same area as the footprint, 
$$
a b=\underline{p}_{\rm side}\wedge\underline{p}_{\rm top}=\pi (D-a)(n_{\rm top}k_{\rm side}-n_{\rm side}k_{\rm top})/\sigma.
$$
Using Equation 1 of the main text this becomes
$$
n_{\rm top}k_{\rm side}-n_{\rm side}k_{\rm top}=1.
$$
Before returning to the same ribbon, all ribbons traversed on the outside of the fiber in the first step has to be visited exactly once (ribbon $1,2,\ldots,N_{\rm step}-1$ of Fig.~1 a of the main text). This can only be done during successive turns around the fiber, and thus $N_{\rm step}=n_{\rm top}$. Also, before returning to the same ribbon, all other ribbons must have been visited exactly once, giving $k_{\rm top}=N_{\rm rib}$, where $N_{\rm rib}$ is the number of ribbons in the fiber. Thus Equation 3 of the main text follows.
\paragraph*{Approximate relationship for wedge angle:}
The curvature tensor for the cylinder surface defined in Figure~1 a of the main text is given in the orthonormal basis  $(\underline{e}_{\theta},\underline{e}_z)$ (see Fig.~S1) as
$$
{\mathbf K}=\left(\begin{array}{cc} 2/(D-a) & 0\\ 0 &0\end{array}\right),
$$
and the unit vector aligned with the ribbons is given by
$$
\underline{e}_{\rm rib}=\sin \gamma\, \underline{e}_\theta+\cos\gamma\,\underline{e}_z.
$$
With this we can calculate the radius of curvature along the ribbons, $R_{\rm rib}$, as
$$
1/R_{\rm rib}=\underline{e}_{\rm rib}\cdot {\mathbf K}\cdot\underline{e}_{\rm rib}=\frac{2\sin^2\gamma}{D-a}.
$$
The wedge angle is now approximated by
$$
\alpha\approx b/R_{\rm rib}=\frac{2b\sin^2\gamma}{D-a},
$$
which by use of Equation 2 in the main text directly gives Equation 4 of the main text.
\paragraph*{Modifications for interdigitated models:} For the interdigitated models Equation 1 and Equation 3 stay the same, while Equation 2, and Equation 4 become
$$
N_{\rm rib} \frac{b}{\cos\gamma}=\pi (D-a)
$$
and
$$
\alpha\approx 2\pi N_{\rm rib}^2 \left(\frac{b}{\pi(D-a)}\right)^3
$$
respectively.
\end{document}